# Plasmonic Antennas as Design Elements for Coherent Ultrafast Nanophotonics

Daan Brinks[1*†], Marta Castro-Lopez[1], Richard Hildner[1‡], Niek F. van Hulst[1,2*].

[1] ICFO – Institut de Ciencies Fotoniques, Mediterranean Technology Park, 08860 Castelldefels (Barcelona), Spain

[2] ICREA – Institucio Catalana de Recerca i Estudis Avancats, 08015 Barcelona, Spain

*Corresponding Authors:
daan.brinks@icfo.es
niek.vanhulst@icfo.es

[†]Present Address:
Department of Chemistry and Chemical Biology
Harvard University
Cambridge MA 02138
daanbrinks@fas.harvard.edu

[‡]Present Address:
Experimentalphysik IV
Universität Bayreuth
95440 Bayreuth, Germany

**Coherent broadband excitation of plasmons brings ultrafast photonics to the nanoscale. However, to fully leverage this potential for ultrafast nanophotonic applications, the capacity to engineer and control the ultrafast response of a plasmonic system at will is crucial. Here, we develop a framework for systematic control and measurement of ultrafast dynamics of near-field hotspots. We show deterministic design of the coherent response of plasmonic antennas at femtosecond timescales. Exploiting the emerging properties of coupled antenna configurations, we use the calibrated antennas to engineer two sought-after applications of ultrafast plasmonics: a subwavelength resolution phase shaper, and an ultrafast hotspot switch. Moreover, we demonstrate that mixing localized resonances of lossy plasmonic particles is the mechanism behind nanoscale coherent control. This simple, reproducible and scalable approach promises to transform ultrafast plasmonics into a straightforward tool for use in fields as diverse as room temperature quantum optics, nanoscale solid state physics and quantum biology.**

**Introduction:** Physics, biology, chemistry and engineering are pushed to ever smaller length scales and ever shorter time scales, lured by the prospect of resolving and utilizing quantum-mechanical phenomena in large, complex, or disordered systems[1,2]. To unlock this regime many efforts have been made to combine ultrafast techniques with plasmonics, which promises optical investigations at deep subwavelength resolutions[3–11]. Examples include polarization control of localization[5,6], measurements of plasmon dephasing[7–9], adiabatic compression of pulses at plasmonic tips[11] and ultrafast excitation of quantum dots in the near field[10]. However, so far a framework for controlling the ultrafast properties of localized fields utilizing plasmonic structures at will has been lacking, as have simple design rules and flexible, scalable and reproducible designs. These are essential prerequisites for ultrafast plasmonics to find widespread use in nanoscale applications.

The Fourier principle states that the ultrafast properties of plasmonic systems can be determined by designing their spectral amplitude and phase profiles. An ideal platform to achieve simplicity, flexibility, scalability, reproducibility and spectral design is that of plasmonic nanoantennas: metallic particles confining far field excitation to near-field hotspots with properties determined by geometry[12–14], material[15] and excitation method[16–18]. A design that has been tested and analyzed particularly rigorously is that of single and coupled bar antennas[12,15,19–24]. These are metal bars of nanometric dimensions, where the width and the height of the bars are typically as small as possible to validate a 1-dimensional approximation. The bars generally exhibit resonant behaviour, where the resonance wavelength can be tuned by the dimensions of the bar[15]. In a first approximation, momentum matching between the plasmon wavevector, the resonance wavevector and the wavevector of the excitation light is required, which determines the wavelengths and polarizations that can excite resonances on particles of a given material with a particular geometry[20,25]. Bringing bars close together creates coupled oscillators, with the coupling strength and thus the resonance splitting being determined by the gap size between the bars[19,22].

The current insights about bar antennas can be distilled down to two important facts: plasmonic antennas confine far-field to near-field, preserving coherence[3,6]; and their resonances exhibit wide bandwidths, which renders them inherently suited for ultrafast processes and coherent control[23,26,27]. However, employment of plasmonic bar antennas as elements for ultrafast, ultralocalized field control runs into the apparent paradox that the response of the antennas needs to be independently characterized, while at the same time their signals should not pollute the signal from any ultrafast coherent process one wishes to influence or catalyse with them. This leads to the added requirement that the plasmonic nanoantenna itself should provide a signal that allows determining the ultrafast dynamics of the fields in its hotspots, but is spectrally shifted from direct coherent higher harmonic signals.

Taking all these conditions into account, gold nanoantennas emerge as the ideal platform for intuitive, ultrafast and ultralocalized field control. In this work we develop a method to measure the ultrafast properties of gold annoantennas by determining the spectral amplitudes and phases in hotspots utilizing a spectrally shifted two photon photoluminescence signal. We characterize the ultrafast properties in the hotspots of single bar antennas and utilize them as building blocks to design coupled plasmonic antennas for two of the most heralded applications of ultrafast plasmonics: a subwavelength resolution phase shaper and an ultrafast spatial hotspot switch.

## Results

**Determining the complex spectrum in a near field hotspot.** For a femtosecond pulse with a spectrum centered at $\omega_0$ and spectral amplitude and phase $E(\Delta)e^{i\phi(\Delta)}$ at $\Delta=\omega-\omega_0$, the Probability of a Two-Photon absorption (PTP) at a particular frequency $\omega_{TP} = 2(\omega_0+\Delta)$ depends on the constructive addition of photons that have a combined energy $\omega_{TP}$. PTP($\Delta$) then depends on the fundamental spectral amplitude and phase as follows[28]

$$PTP\left(\Delta\right) \propto \left| \int \left| E\left(\Delta+\Omega\right) \right| \left| E\left(\Delta-\Omega\right) \right| e^{i\left[\varphi\left(\Delta+\Omega\right)+\varphi\left(\Delta-\Omega\right)\right]} d\Omega \right|^2 \qquad (1.1)$$

where the integration variable $\Omega$ is a detuning from the central frequency $\omega_0$ similar to $\Delta$.

Assuming continuous phases, $\varphi\left(\Delta+\Omega\right)+\varphi\left(\Delta-\Omega\right)$ can be expanded into a Taylor series where the second derivative of the phase is the first term that does not integrate out in equation (1.1). This consideration gives us a condition to measure the spectral phase added to an ultrafast hotspot by an antenna: the signal at a particular $\omega_{PTP}=2(\omega_0+\Delta)$ will be maximum if $\varphi''(\Delta)=0$.

To measure this antenna phase in the hotspot, we expand $\varphi''(\Delta)$ into a known phase added in a pulse shaper and an unknown antenna phase, $\varphi''(\Delta)=\varphi_{shaper}''(\Delta)+\varphi_{dispersion}''(\Delta)$, and we thus need to solve

$$\varphi_{shaper}''(\Delta)+\varphi_{dispersion}''(\Delta)=0 \qquad (1.2)$$

Analogous to the Multiphoton Intrapulse Interference Phase Scan (MIIPS) method[28–36], we add a series of cosinusoidal phase functions to a 120 nm, 15 fs pulse (fig. 1a) in a pulseshaper: $\varphi_{shaper}(\omega)=\alpha\cos[(\omega-\omega_0)\beta+\delta]$ (fig 1b). $\alpha$ is the amplitude of the phase function and determines the resolution with which the phase can be retrieved; $2\pi/\beta$ is the periodicity of the phase function in frequency space, determining the bandwidth over which the phase can be measured; $\delta$ is a phase offset. Scanning the phase offset $\delta$ (fig. 1b) samples the phase the antenna adds to the pulse (fig. 1c). This allows us to record an antenna "signature" (fig. 1d): a Two-Photon Photoluminescence (TPPL) signal as a function of $\delta$ that traces the PTP and thus contains information about the phase the antenna adds to the field in the antenna hotspot. This method has the added value that no amplitude modulation is applied to the laser pulse by the shaper; the approach allows for measurements specifically sensitive to spectral phase, free of spatio-temporal coupling.[37]

The cosine function has two well-defined points per period where $\varphi''_{shaper}(\Delta)=0$, namely in the zero crossings. These are scanned through the spectrum by varying the frequency offset $\delta$. (fig 1b). Appropriate tuning of $\beta$ ensures that for a variation of $\delta$ from 0 to $2\pi$, both zero-crossings traverse the spectrum one after the other. To understand the measurement, it is important to compare two cases: without and with antenna dispersion.

Without dispersion, the zero crossings of the cosine in $\phi_{shaper}(\Delta)$ determine the frequencies in the spectrum that dominate the PTP, since those frequencies are the only ones that are phasematched. The TPPL signal we measure depends on the total probability of generating a two-photon excitation in gold. This is proportional to the spectrally integrated PTP and is in the zero-dispersion case therefore dominated by the phasematched frequency, which is directly given by δ through $\phi"_{shaper}(\Delta) = 0 \rightarrow \omega - \omega_0 = \dfrac{\delta \pm \pi/2}{\beta}$. When dominated by the shaper phase, the TPPL response as a function of δ therefore traces the fundamental spectrum twice (fig. 2).

With antenna dispersion, the relative contribution of the affected wavelengths to the PTP will change, which will be reflected in the TPPL response as a function of δ (Fig 1c-d). This trace now therefore provides a signature of the antenna dispersion: For each δ, the integrated PTP will be dominated by the frequencies for which the total dispersion happens to be minimal. The shape of the signature will therefore deviate from the fundamental spectrum. Applying prior knowledge about the amplitude of the fundamental spectrum therefore allows a fit to the signature which allows retrieving the spectral phase.

When it is taken into account that $\varphi_{shaper}\left(\omega, 0 < \delta < \pi\right) = -\varphi_{shaper}\left(\omega, \pi < \delta < 2\pi\right)$, it can be seen that two measurements are performed, one with a "positive phase" and one with a "negative phase". This means we are obtaining a quasi-2D solution to the 2D problem of measuring an unknown amplitude and phase; in other words, comparing the two halfs of the antenna signature allows disentangling the amplitude and phase modulation the antenna adds to the field in the hotspot.

**The building blocks: single bar antennas.** The key to using nanoantennas in plasmonic designs is a systematic measurement of the complex spectral modulations that antennas of increasing lengths add to the hotspot-field. Figure 3a presents the antenna signatures of single bar antennas with increasing lengths from 480 to 640 nm, recorded with laser pulses of 15 fs width (corresponding to a spectral bandwidth of 120 nm at 776 nm, see fig. 1a and SI). The parameters for the cosinusoidal phase are $\alpha = 0.9\pi$ and $\beta = 14\ fs$. An intuitive understanding of these data can be gained by considering that the highest TPPL signal occurs when the shifting cosinusoidal phase $\varphi_{shaper}$ compensates for the spectral phase of the antenna resonance best. The particular shape of the band of maximum intensity that starts at $\delta = 0$ rad at 480 nm rods and curves towards $\delta = \pi$ for longer rods (fig. 3a) can thus be interpreted as the phase jump that the excitation light at carrier wavelength 776 nm experiences when it drives the antenna above or below the resonance frequency. Recording the signature between $0 < \delta < \pi$ and $\pi < \delta < 2\pi$ allows us to differentiate between the effect the antenna has on the spectral amplitude and on the spectral phase of the field: this can clearly be seen in e.g. the case of a 560 nm antenna: when δ is ~π/2 the cosine function compensates best for the dispersion of the antenna, and the combination with a good overlap of the excitation spectrum with the resonance of the antenna provides for a high two-photon signal; in contrast, when δ is ~3π/2, the overlap between the excitation spectrum and the antenna resonance is equally good but the cosine function adds to the antenna dispersion, and the two-photon signal is very low.

To retrieve the spectral resonance profiles of the antennas, the measured signatures are compared to fits based on Lorentzian resonances that are given by

$$L(\omega) \propto \frac{1}{\omega_{cent} - \omega - i\gamma} + \frac{1}{\omega_{cent} + \omega + i\gamma}$$ (fig. 3b). Here $\omega_{cent}$ is the central frequency of the

resonance, 2γ is the full width at half maximum (FWHM) of the imaginary (i.e. absorptive)

part of the resonance and the antenna phase is defined as $\varphi_{antenna}(\omega) = \arg[L(\omega)]$. The best fits are obtained if the width of the Lorentzian is kept at $2\gamma$=0.13±0.02 rad fs$^{-1}$ (or $\Delta\lambda$= 41±6 nm at 776 nm central wavelength) while the resonance wavelength is varied from 640 nm for antennas of 480 nm length to 940 nm for antennas of 640 nm length (corresponding respectively to $\omega_{cent}$=2.95 rad fs$^{-1}$ and 2 rad fs$^{-1}$).

**Plasmonic near-field phaseshapers.** The first systematic measurement of the complex spectrum of antennas as a function of length over a wide range (fig. 3) allows us to create hotspots with tailored ultrafast responses by coupling antennas. The field in each hotspot in such a system is influenced by the resonances of the building blocks, modified by the dispersion and absorption of the coupled system. An application that suggests itself immediately is the creation of a subwavelength resolution phaseshaper: a plasmonic structure that imparts a predefined phaseshape on the field in a hotspot for use in high resolution, low volume nonlinear experiments. This device can be engineered by combining single antenna bars as building blocks into a linearly coupled asymmetric antenna, separated by a nanometer sized gap.

This concept of engineering ultrafast nanophotonic systems is demonstrated in figure 4 where we present the retrieved amplitude-phase profiles of the hotspots on a single bar antenna of 530 nm (fig. 4a) and 620 nm length (fig. 4b), as well as the 530 nm (fig. 4c), resp. the 620 nm side (fig. 4d) of a coupled antenna with a gap of ~30 nm. The comparison of the signatures of antennas of the same length, either coupled to another antenna or not, reveals distinct differences in the amplitude- and phase-profiles in their hotspots (compare figs. 4a and 4c; figs. 4b and 4d).

Insight into the mechanism that determines the ultrafast properties in each hotspot is gained by fits to the measured signatures. Figure 4 shows the data with the best fit (left) and the corresponding reconstructed resonance profile (right). The best fits for the signatures were obtained by significantly broadening the single antenna resonances from 0.13±0.02 to 0.26±0.06 rad fs$^{-1}$ for coupled antennas, indicating that the total dispersion of the combined antenna determines the width of the resonances in each profile. The losses in the metal, giving rise to the broadening, are a vital property of the design, because they largely determine the range of the spatial mode associated with the resonance and therefore its strength in a particular hotspot. In this way, a plasmonic nanoantenna can be employed as a subwavelength resolution phase shaper, as shown in fig. 4 (c,d, right): the phase the antenna imprints on the hotspot-sized field can directly be tuned by adapting the strength and separability of the resonances of the constituent antennas in the hotspot.

**Ultrafast plasmonic switch**. The reconstructed phase-profiles in the hotspots give insight into the phase-sensitive properties of hotspots on coupled nanoantennas, and more specifically the role of the spatial range of resonant modes. This now allows us to redefine coherent control in coupled plasmonic systems in an intuitive way as addressing spectrally separable resonances in a particular hotspot with the appropriate strength and timing to achieve the desired temporal field structure. Hotspot switching can then be realized by simply addressing localized resonances at different times in an excitation pulse through application of a spectral phase. To demonstrate this concept, we engineer a simple asymmetric coupled antenna to function as a sub-100 fs spatial hotspot switch.

Antennas with lengths of 530 nm and 620 nm are chosen as suitable building blocks because they feature resonances on the high-, resp. low frequency side of our excitation spectrum. The coupled antenna is excited with a negatively chirped pulse (-500 fs$^2$ chirp). Using the asymmetry in the resonances of the different hotspots (as shown in fig. 4c,d, right), this creates a concentration of intensity on the short antenna bar in the first half of the pulse, switching to the long antenna bar in the second half of the pulse. The dynamics of the field is resolved in a pump-probe experiment by employing a time-delayed, Fourier-limited probe pulse (fig. 5a).

The ratio between the TPPL intensity from the hotspots on the 530 resp. 620 nm side of the coupled antenna is depicted in fig. 5b (right, gray points) as a function of pump-probe delay. The intensity switches from the 530 nm side to the 620 nm side between delays of -50 fs and +35 fs with appreciable contrast ( ~2.8). The comparison between experiments and theory based on the resonance structure of the hotspots and the chirp in the excitation pulse (fig. 5b, right, solid black line) shows a good agreement between the actual hotspot switching of the antenna and the theoretical behaviour for which the antenna was designed ($R^2$=0.92). The ultrafast switching can directly be visualized by imaging the antennas with fixed pump-probe delay at -50 fs (fig. 5b, top left), resp. 35 fs (bottom left). This demonstrates the utility of a coupled antenna for hotspot switching within 85 fs, applying only quadratic chirp in the excitation pulse.

**Discussion:** Our scheme shows the potential for subwavelength coherent control and spectrally selective ultrafast microscopy without the need for a phaseshaper; one can design a plasmonic structure that will have the required spectral phase for a desired multiphoton excitation process in the hotspot of interest, allowing for the integration of ultrafast and phase sensitive measurements in chip-based constructs. In each hotspot the total spectral polarization, phase and amplitude can be calculated by weighted vectorial addition of the resonances of each bar. Coupled antenna bars are thus turned into an intuitive tool to control the ultrafast characteristics of specific subwavelength excitation spots.

It is tempting to discuss our results in terms of coupling strengths between oscillators. While changes in the phase profiles in off-center hotspots do prove coupling between the bar antennas, a treatment in these terms foregoes the distinguishing feature of this work: we go beyond integral plasmonic constructs as coupled point oscillators and explicitly look at the contribution of resonances in individual hotspots, i.e. on a "sub-resonator" level. It has recently been proven that the behaviour of coupled plasmonic oscillators is classical down to the smallest gap sizes that can be created to date with top down fabrication[38]; in this light, it makes more sense to treat the dynamics in each hotspot separately in terms of the strength of the resonances contributing to the particular amplitude-phase profile in the hotspot.

Another way of looking at this is that coupling antenna bars creates artificial, plasmonic molecules, where the different modes correspond to different molecular resonances[39,40]. While the analogy is not perfect, it serves to illustrate an important point: understanding of not only the linear resonance of a "plasmonic molecule" (the absorption spectrum) but the nonlinear properties as well (the spectral phase) allows for designing plasmonic structures that selectively excite a real molecule with an analogous nonlinear absorption spectrum within a subwavelength sized hotspot. This provides an intuitive way to create plasmonic structures to selectively enhance the excitation of a particular species of molecule with high spatial resolution in a mixture or supramolecular construct, using not only the direct field enhancement but also phasematching the plasmonic and molecular resonances.

Pump-probe experiments resolved at a nanometric level are easily performed with our approach: the switching time in the ultrafast plasmonic switch can be scanned by changing the chirp in the excitation pulse, simply through adding dielectric material in the beam path. Bandwidths, switching times, localization contrast and phase gradients are all governed by the prominence of the resonances in each hotspot; material properties, gap sizes and relative orientations provide a wide range of parameters to tune the resonance-strength and -separability in the hotspots.

Investigations at ultrafast timescales are being rapidly embraced in fields as diverse as semiconductor physics, cell biology and quantum optics. What unites these areas is the transition from research on phenomena in isolated or uniform systems (molecular jets, ultracold atoms, solutions) to nanoscale components interconnected in larger systems (optoelectronic switches, supramolecular complexes, cells, NV-center networks). For this paradigm shift, femtosecond time resolution needs to be combined with high spatial selectivity and complete control of the time-evolution of fields with nanometer precision, as we have demonstrated here.

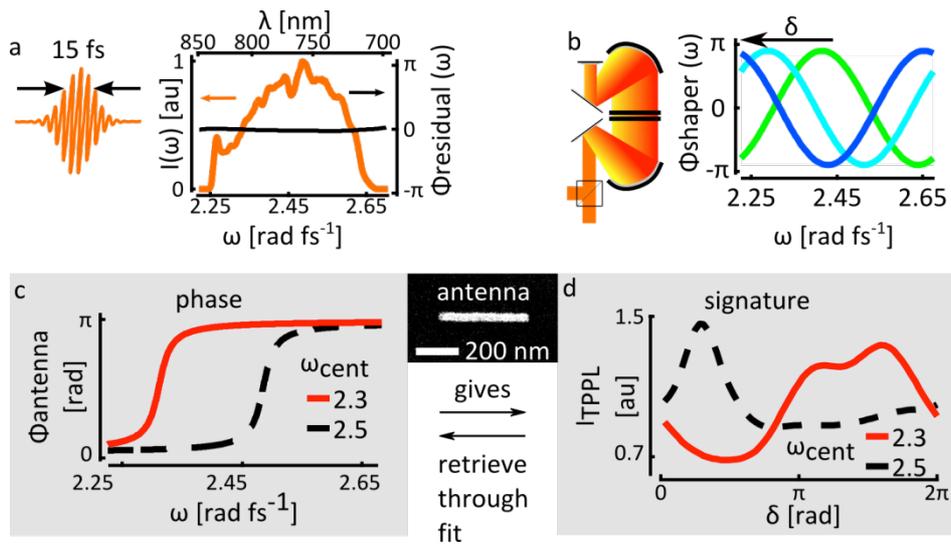

**Fig. 1: Characterization of spectral amplitude and phase in nano-antenna hotspots.** a) A gold nanoantenna (inset) is excited with a Fourier limited 15 fs pulse with depicted spectrum and residual phase $\Phi_{residual}$ flat within 0.05)π rad. b) A series of cosinusoidal phases with shifting offset δ is added to the pulse in double pass 4f pulse shaper. c) The antenna resonance adds a particular phase to the field in a hotspot. d) The shifting cosinusoidal phase samples the antenna phase, which gives an antenna signature that can be fitted to retrieve the phase profile (see also Fig. 2).

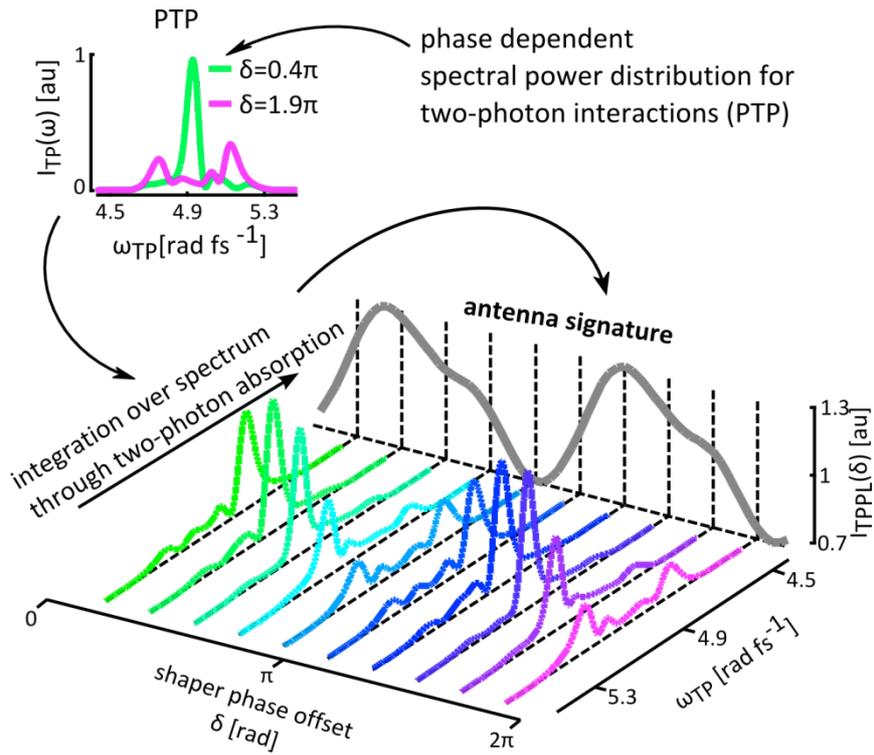

**Fig. 2. Mapping the antenna phase in a TPPL signature.** Inset: The PTP for each δ is dominated by the frequency $\omega_{TP}$ for which the second derivative of the total spectral phase $\varphi''(\Delta) = 0$: $I_{TP}(\omega)$ is sharply maximum for this frequency. Main figure: Spectrally integrating the PTP for each δ, through two-photon absorption, traces this characteristic behavior into a phase signature. Without antenna dispersion, this signature tracks the shape of the fundamental spectrum twice (grey line). With antenna dispersion, integrating the PTP for each δ provides an antenna signature that differs from the zero dispersion case in a characteristic way and therefore provides a direct measurement of the spectral phase that the antenna adds to the field in the hotspot (see also fig. 1c,d).

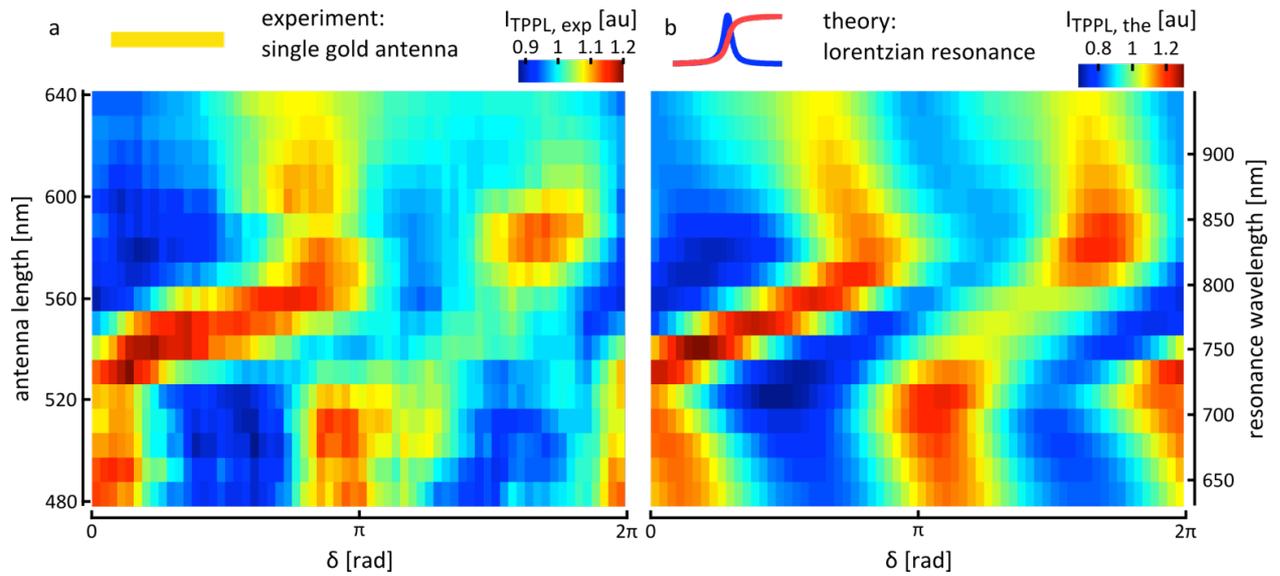

**Fig. 3. Coherent ultrafast dynamics of single nanoantenna hotspots.** a) Antenna signatures of gold nano-antennas of increasing lengths. The signatures measure the spectral phase and therefore, through the Fourier transform, the ultrafast dynamics of the hotspot. b) Corresponding signatures calculated for Lorentzian resonances at different wavelengths with a width of 0.13±0.02 rad fs$^{-1}$ (or 41±6 nm at 776 nm).

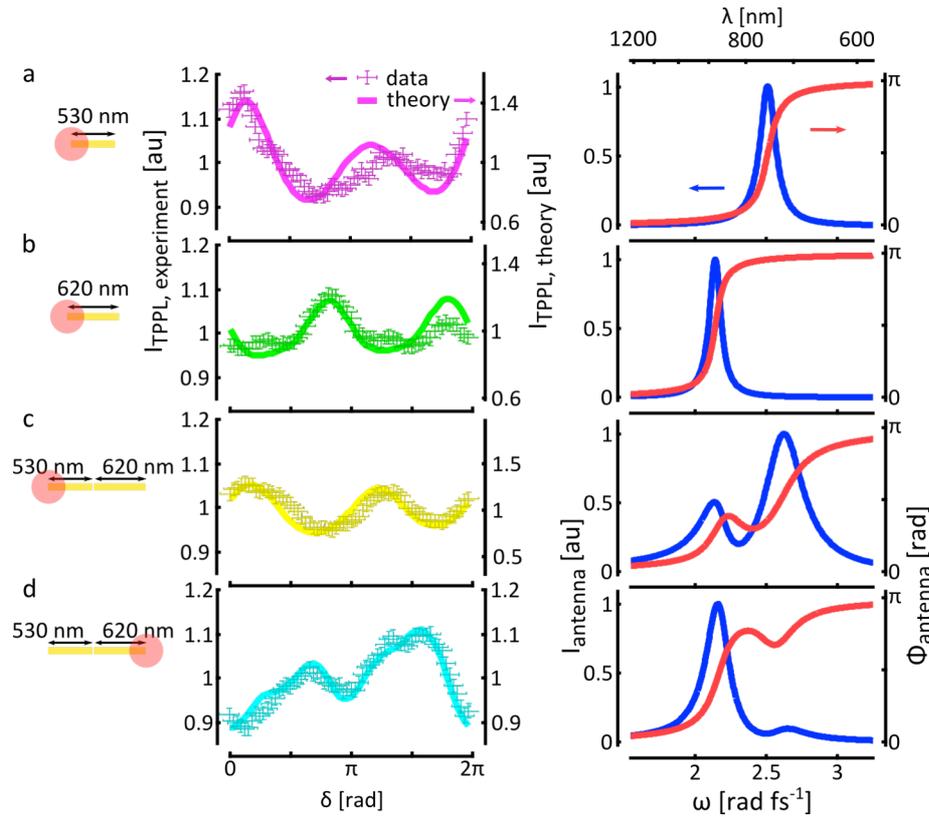

**Fig. 4. A subwavelength resolution plasmonic phaseshaper.** a) Antenna signature measured in a hotpot on a 530 nm single antenna (left) and the resonance profile retrieved from the fit (right). b-d) Corresponding signatures (left) and resonance profiles (right) for a 620 nm single antenna and different positions on a coupled 530-620 nm asymmetric antenna, demonstrating the engineering of spectral phase in antenna hotspots by mixing resonances with different strengths.

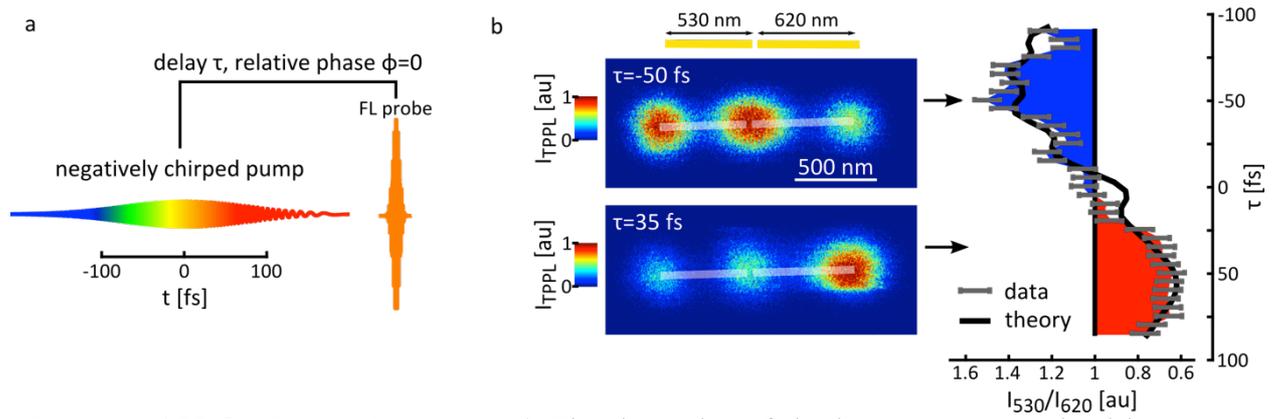

**Fig. 5. A sub-100 fs plasmonic switch.** a) The dynamics of the hotspots are resolved in a pump-probe experiment. b) Between a pump-probe delay of -50 fs and 35 fs the luminescence intensity switches from the 530 nm antenna bar to the 620 nm antenna bar with a contrast ratio ~ 2.8 (right). Imaging the antenna with these delays fixed, demonstrates the sub-100 fs plasmonic switching (left).

**Supplementary Materials:**

**Method Section**

The output of a broadband Ti:Sapphire laser (Menlo-Systems Octavius 85 M) was expanded and collimated with a set of spherical mirrors and fed into a double pass 4f-pulse shaper based on an in-lab modified MIIPS-box. A 120 nm spectral band centered around 776 nm of the laser output was dispersed with a grating and focused on a 640 pixel double layer spatial light modulator (SLM). The light was reflected back through the SLM, recollimated and recombined, caught on an end-mirror and reflected back through the entire shaper. The shaper output was separated from the input with a polarizing beam splitter and directed through a Glan-Taylor polarizer to clean the amplitude modulation. The beam was subsequently spatially filtered using a telescope with a 100 μm pinhole.

The experiments were performed on a modified confocal microscope (Zeiss Axiovert). The shaped pulses were led into the back aperture of the microscope and reflected into the sample with a short-pass dichroic mirror (Semrock SDi-01-670). The pulses were focused into the sample plane with a 1.3 NA objective (Zeiss Fluar). The two-photon photoluminescence (TPPL) of the antennas was collected through the same objective, separated from the excitation light by the dichroic mirror and two short pass filters (Semrock FF01-720SP-25 and FF01-660SP-25) and focused on an APD (Perkin Elmer SPCM-AQRH-16).

Pulse calibration was done via the MIIPS method with a micrometer sized BBO-crystal plate in the sample plane. The second harmonic (SH)-signal of the crystal plate was collected in transmission by an optical fiber and focused onto an imaging spectrograph with a sensitive charge-coupled device (CCD) camera (Andor SR-163 with camera Andor DV437-BV). The calibration resulted in a compression mask: a phase added in the shaper to ensure a Fourier limited starting pulse in the sample plane; the pulse shapes utilized in the experiment were added on top of this. For calibration and measurement integration times of at least 20 seconds were used to ensure phase stability of the SH and TPPL signals.

The experiments were performed on gold nanoantennas of variable length, 50 nm width and 20 nm thickness. The sample was fabricated on 10 nm ITO on glass by e-beam lithography, thermal gold evaporation and lift-off, according to a previously described procedure [15]. On the same sample, matrices with repetitions of single antennas with varying lengths and coupled antennas with different ratios between bar lengths were alternated.

The sample was excited with an 85 MHz train of pulses centered at 776 nm with the spectral amplitude and phase as shown in fig. 1a. The time-averaged power in the sample plane after spatial filtering and application of additional neutral density filters was $3.3 \cdot 10^{-6}$ W, corresponding to a flux of ~2.8kW/cm$^2$.

The antennas were excited with linear polarization along the antenna axis; in combination with the linear antennas this served to avoid any polarization effects.

At the start of each experiment the sample was scanned through the focus of the microscope objective with a piezo scanner (Mad City Labs Nanoview/M 100-3), yielding TPPL images of the matrices of antennas with each hotspot lighting up. The experiments were performed by positioning one of the hotspots in the focus with the piezo scanner. Subsequently the phase shape of the pulse was changed, and the corresponding TPPL response of the antenna was recorded. Typically, all phase shapes were cycled with a 1 s integration time each. The measurement for every phase shape was alternated with an equally long measurement with a Fourier limited reference pulse. This ensures long total integration times for each signal point, and at same time provides the time resolution and reference signal to monitor signal changes unrelated to the experiment. Depending on the antenna and its resonance, the signal was between 2.5 and 6.5 kcts/s. For the data from figure 3 and 4, this cycle was repeated between 20 and 30 times; each cycle consisted of 64 measurements of different phase shapes, alternated with 64 reference measurements. The signal traces are based on total amounts of counts between 50 and 200 kcts per point. For the data in figure 5, the cycle consisted of 36 delay measurements, interleaved with 36 reference measurements. This cycle was repeated 5 times; the signal traces are based on a total amount of counts between 12 and 35 kcts per point. For the phase determination we employ a least squares fit in a basis of Lorentzian resonances where the number of resonances, their central wavelengths, their widths and their relative weights are free parameters.

The signal error bars in all graphs are the magnitude of deviations of the 64 reference measurements to their average. The errorbars for figure 4 also hold for figure 3 and are 1.5%; note that in figure 3, although the measurements were performed on individual antennas and the stated error holds for that, the graph shows the average signature of three antennas to enhance the signal to noise. The δ-errorbars in figure 4 reflect the calibration accuracy for the Fourier limited pulse: the residual phase after compression was flat to within $0.1\pi$ radian. This error practically falls away in the significantly larger phase added to the pulse in figure 5: -500 fs$^2$chirp gives a quadratic phasefunction with a range of $5\pi$ radians throughout the pulse spectrum; the error resulting from the uncertainty in compression amounts to a 2% uncertainty in delay, i.e. maximum ±2 fs at 100 fs delay. The signal error bars in figure 5 are also determined by the deviation of the reference measurements to their average and are in the range of 6% due to the shorter integration time and the division between 2 traces.

**The pump probe experiment**

For the pump probe experiment the intensity of the TPPL signal in the hotspots is recorded as a function of the delay between the chirped pump pulse and a Fourier limited probe pulse, with both pulses having the same spectrum. This double pulse was created in a pulse shaper. In order to avoid fast interference fringes, the probe pulse was phaselocked to the central frequency of the band it was overlapping in the chirped pump pulse at each delay (fig. 5a main manuscript). The total probability for a two-photon-excitation in each hotspot depends on the initial spectrum of the pulse (fig. 1a), the chirp added in the pump pulse (-500 fs$^2$), the amplitude and phase profiles in the hotspots (fig. 4 main manuscript), and the delay and phase between the pump and the probe pulse (fig. 5a main manuscript). The theoretical curve was calculated by integrating the PTP for each pump-probe pair in both hotspots, and dividing the two values.

**TPPL signals from gold**

For our purpose, non-crystalline gold antennas are attractive because they show very little direct Second harmonic signal, which means that they will not pollute the coherent signals one tries to create, enhance, influence or probe with the antenna. They do however provide a large Two-Photon photoluminescence signal (TPPL). TPPL has been used as a characterization tool for hotspots from the earliest works on plasmonic antennas[12,16,41]. This is a great way to obtain nonlinear information about eh ultrafast dynamics in hotspots without interfering with future measuerments one wises to do with the antenna structures. The good qualitative agreement, shown in figure 3 of the main manuscript, in the trend between the signatures measured on the single antenna bars and calculated for the reconstructed resonances demonstrates that viable phase information can be extracted from the TPPL measurements of the antenna signatures. However, the contrast in the measured signatures is markedly less than that in the calculated signatures. This can have several reasons. First, liquid crystal based phase shapers work under the approximation that a particular pixel in the liquid crystal (LC) corresponds to a particular wavelength in the spectrum. In reality however, each pixel will receive a band of wavelengths determined by the pixel size and the NA of the optics focusing the light on the LC. Hence, each wavelength will be focused on a band of pixels determined by the same parameters. As a consequence of this, a phase function added in the shaper $\phi(\omega)$ transforms to the actual phasefunction $\varphi(\omega)$ given by $\varphi(\omega) = \int \phi(\Omega) g(\omega - \Omega) d\Omega$, where ideally $g(\omega) = \delta(\omega)$, the Dirac delta function, but in reality is a function with a finite width. Effectively this means that amplitude α of the applied function will become slightly lower than in theory, limiting the contrast in the recorded signature.

Another source of limited contrast is the phase noise in the excitation pulse. When the residual phase in the spectrum fluctuates, it is possible to obtain a PTP with the correct shape by integrating long enough, but the total power in the integrated PTP will be lower and the effective amplitude α of the added phase function will go down with the width of the noise band.

Finally, an interesting possibility was posited by Biagoni et al [42,43] that an intermediate state in the gold is involved in the generation of TPPL. The exact dependency of a TPPL signal on this state is elusive; it shows varying behaviour depending on pulse widths, peak powers, central wavelengths, etc. The variation in the signatures clearly shows a phase dependence of the TPPL signal, but a short lived intermediate level is a likely candidate in accounting for the loss in contrast in the measured signal compared to theory. However, a full treatment of its causes and influences goes beyond the scope of this paper.